\documentclass[letterpaper]{article} 
\usepackage{aaai24}  
\usepackage{times}  
\usepackage{helvet}  
\usepackage{courier}  
\usepackage[hyphens]{url}  
\usepackage{graphicx} 
\usepackage{natbib}  
\usepackage{caption} 
\usepackage{algorithm}
\usepackage{algorithmic}
\usepackage{xcolor}

\NewDocumentCommand{\heng}
{ mO{} }{\textcolor{red}{\textsuperscript{\textit{Heng}}\textsf{\textbf{\small[#1]}}}}

\usepackage{newfloat}
\usepackage{listings}
\DeclareCaptionStyle{ruled}{labelfont=normalfont,labelsep=colon,strut=off} 
\floatstyle{ruled}
\newfloat{listing}{tb}{lst}{}
\floatname{listing}{Listing}
\title{InfoPattern: Unveiling Information Propagation Patterns in Social Media}
\author{
    Chi Han\textsuperscript{\rm 1}\equalcontrib, Manling Li\textsuperscript{\rm 3}\equalcontrib, 
    Jialiang Xu\textsuperscript{\rm 2}\equalcontrib, Hanning Zhang\textsuperscript{\rm 1}\equalcontrib, Tarek Abdelzaher\textsuperscript{\rm 1}, Heng Ji\textsuperscript{\rm 1}
}
\affiliations{
    \textsuperscript{\rm 1}UIUC  \quad   \textsuperscript{\rm 2}Stanford    \quad
    \textsuperscript{\rm 3}Northwestern \\

    {\small \texttt{\{chihan3,hanning5,zaher,hengji\}@illinois.edu, xjl@stanford.edu, manling.li@northwestern.edu}}
}

\usepackage{bibentry}

\def\signed #1{{\leavevmode\unskip\nobreak\hfil\penalty50\hskip2em
  \hbox{}\nobreak\hfil(#1)%
  \parfillskip=0pt \finalhyphendemerits=0 \endgraf}}

\newsavebox\mybox
\newenvironment{aquote}[1]
  {\savebox\mybox{#1}\begin{quote}}
  {\signed{\usebox\mybox}\end{quote}}
  
\usepackage{amsfonts}

\begin{document}

\maketitle

\begin{abstract}

Social media play a significant role in shaping public opinion and influencing ideological communities through information propagation. 
Our demo \textit{InfoPattern} centers on the interplay between language and human ideology.  
The demo\footnote{\textbf{Code:} \url{https://github.com/blender-nlp/InfoPattern}} is capable of: (1) red teaming to simulate adversary responses from opposite ideology communities; (2) stance detection to identify the underlying political sentiments in each message; (3) information propagation graph discovery to reveal the evolution of claims across various communities over time. 
\footnote{\textbf{Live Demo:} \url{https://incas.csl.illinois.edu/blender/About}}
\end{abstract}

\section{Introduction}


\begin{aquote}{Sir Winston Churchill}
\textit{``We shape our buildings; thereafter they shape us.''}
\end{aquote}

Language, as the core of human communication, carries more than just information—it embodies cultural, demographical, and social contexts. 
Moreover, the deliberate framing of information, or propaganda, can manipulate language to serve specific ideological agendas, revealing a reciprocal relationship between language and human intention \citep{salmon1989manufacturing}. 
As an example, the dissemination of information has the potential to sway presidential elections \citep{doi:10.1126/science.aaw8243}. 
Similarly, during the COVID-19 pandemic, claims from different communities significantly influenced public trust in vaccines \citep{BURKI2020e504}. In this paper, we present a novel InfoPattern demo to characterize the intent behind messages and further uncover the information propagation patterns in social networks.


%
Recent efforts have offered insights into stance detection~\citep{10.1111/jcom.12077,10.1007/978-981-15-0029-9_7,DRUS2019707,ALDAYEL2021102597} and response prediction in information spreading~\cite{tan2023botpercent,sun2023measuring,app11083697}, 
but with a focus on local reactions, lacking in global analysis over the information propagation for multiple communities. In this demo, we aim to delve into these intricate dynamics, shedding light on how language generation can be conditioned on the underlying intention and, in turn, influence claims from different ideology groups. 

We propose our framework to contain three functions to simulate such interplay: (1) \textbf{Red Teaming for adversarial ideology communities} to simulate the framing strategies from opposing ideologies \textsc{Left}, \textsc{Neutral}, and \textsc{Right}, by learning a controlling parameter to steer the generation based on LM-Switch~\cite{han2023lm}; 
(2) \textbf{Stance Detection} to classify any text sentence to the underlying political sentiments based on approximating LM-Switch embedding space. 
(3) \textbf{Information Propagation Pathway Discovery} to reveal how claims (core idea) evolve across different communities over time, by automatically detecting claims and learning the claim temporal transition matrix. 

\begin{figure}[t]
    \centering
    \includegraphics[width=0.4\textwidth]{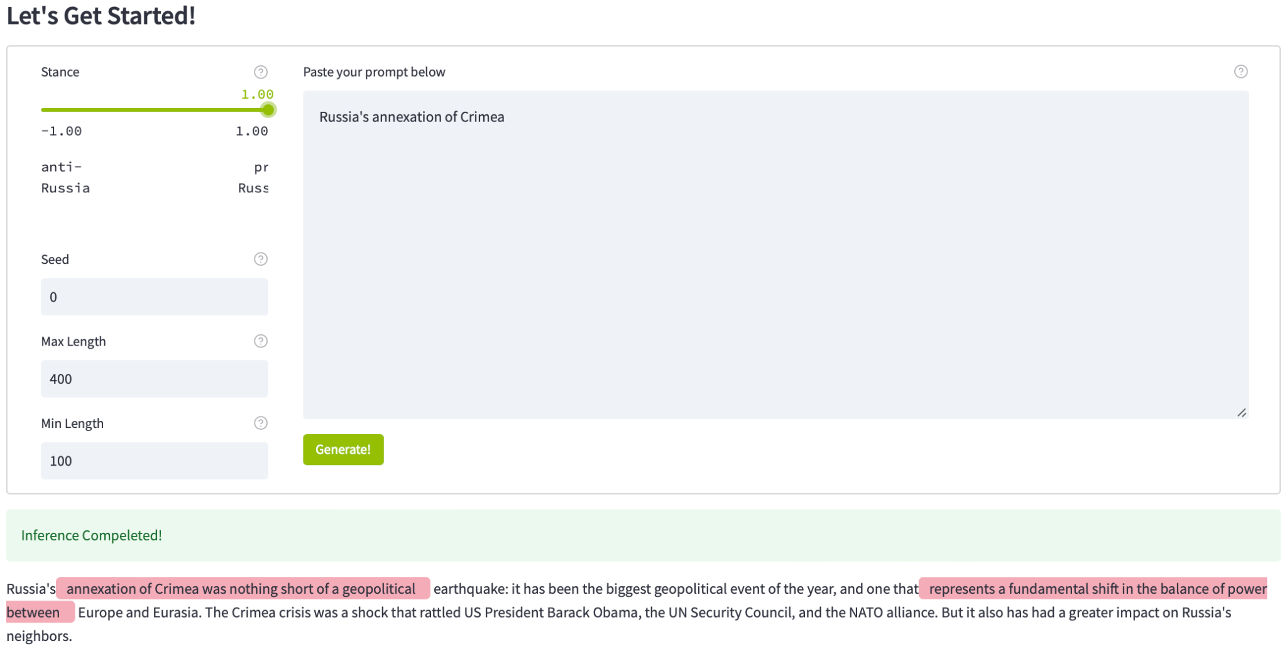}
    \caption{The interface of Red Teaming.}
    \label{generation}
    \vspace{-10pt}
\end{figure}

\begin{figure}[t]
    \centering
    \includegraphics[width=0.4\textwidth]{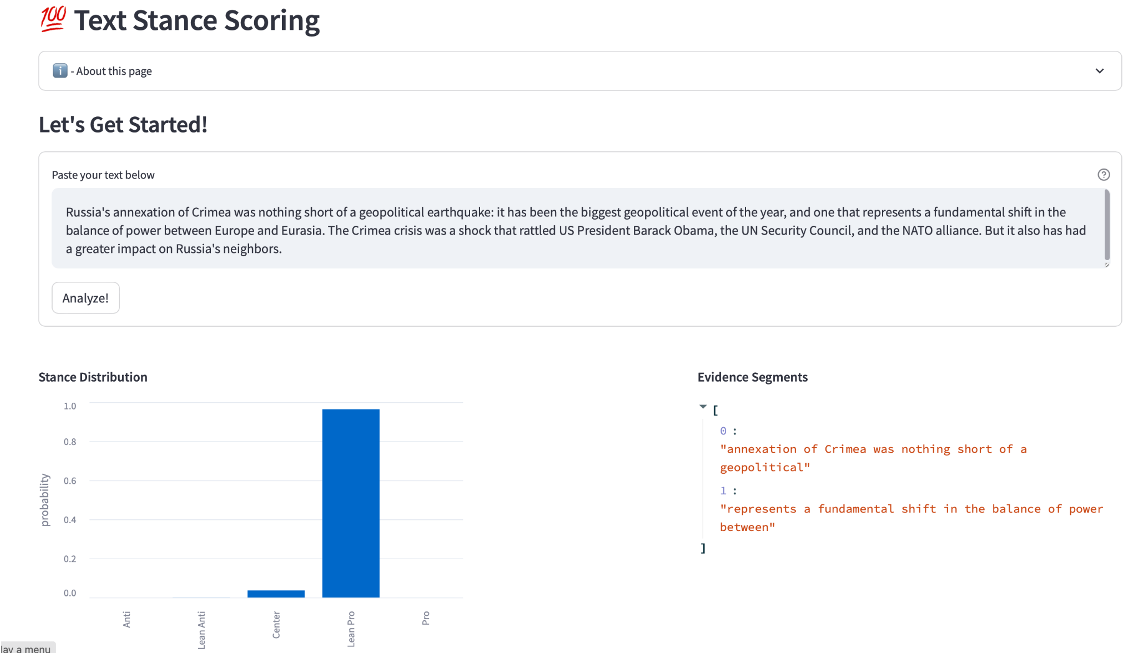}
    \caption{The interface of Stance Scoring.}
    \label{stance}
    \vspace{-10pt}
\end{figure}

\begin{figure*}[t]
    \centering
    \includegraphics[width=1\textwidth]{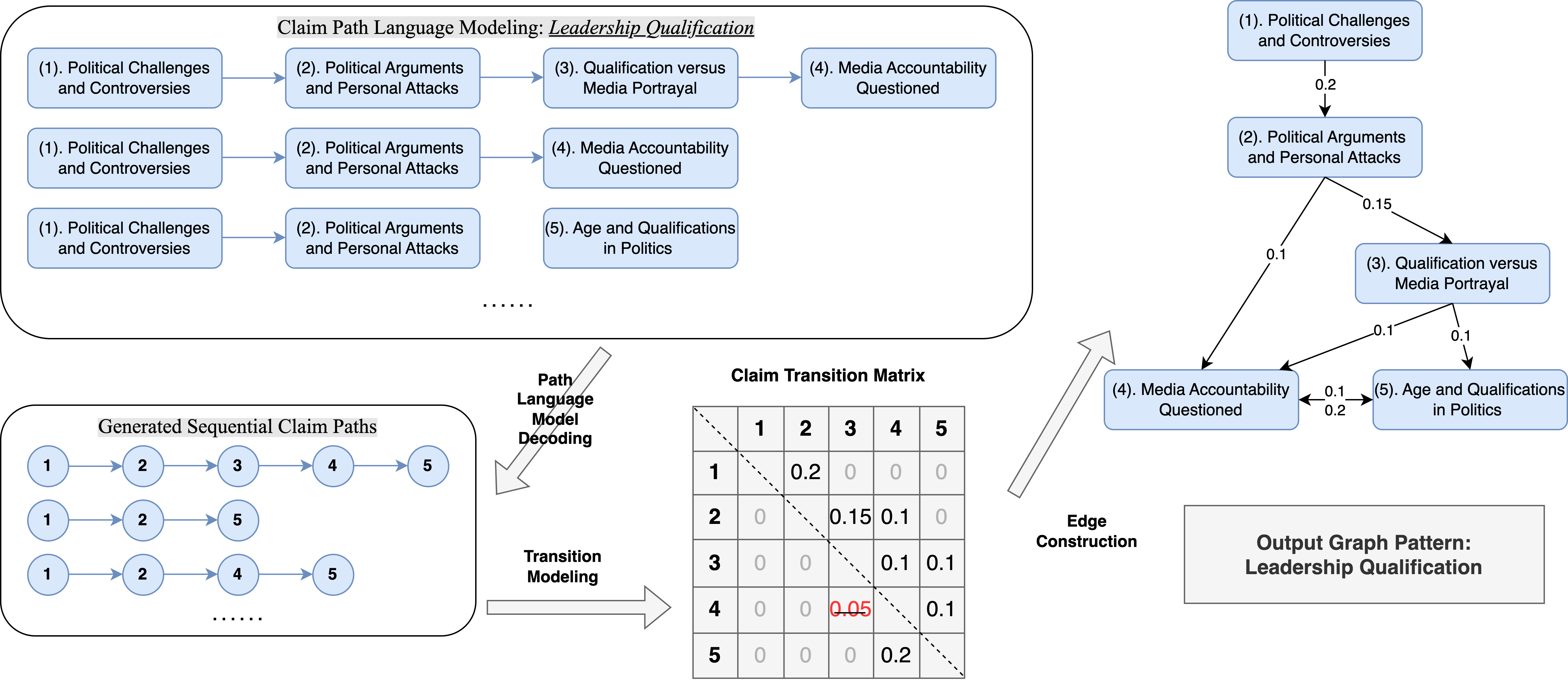}
    \caption{The model architecture of Information Propagation Path Discovery.}
    \label{diagram}
    \vspace{-10pt}
\end{figure*}

\section{Red Teaming and Stance Scoring}
\textbf{Dataset Collection.}
We leverage MultiAgencyNews\footnote{\url{https://github.com/blender-nlp/MultiAgencyNews}} which consists of news articles on the same event with contrasting political stances. 
Each news article is associated with one of the five possible stance labels (\textsc{Left}, \textsc{LeanLeft}, \textsc{Neutral}, \textsc{LeanRight}, and \textsc{Right}).

\noindent
\textbf{Red Teaming.} We leverage LM-Switch~\cite{han2023lm} to learn to steer the language model GPT-J-6B~\cite{gpt-j} based on political spectrum controlling parameter $\varepsilon$ spanning across \textsc{Left} ($\varepsilon=-1$), \textsc{Neutral} ($\varepsilon=0$), and \textsc{Right} ($\varepsilon=+1$). The mechanism underlying LM-Switch is to learn a linear transformation on word embeddings to steer the generation distribution, which implicitly identifies the embedding subspaces most associated with political stances.

\noindent
\textbf{Stance Scoring.} A learned LM-Switch can be naturally applied to detect the stances of \textsc{Left}, 
\textsc{LeanToLeft}, \textsc{Neutral}, \textsc{LeanToRight}, and \textsc{Right}. Specifically, we search within LM-Switch for the stance with the highest likelihood.


\section{Information Propagation Pathway Discovery}

\begin{figure}[t]
    \centering
    \includegraphics[width=0.4\textwidth]{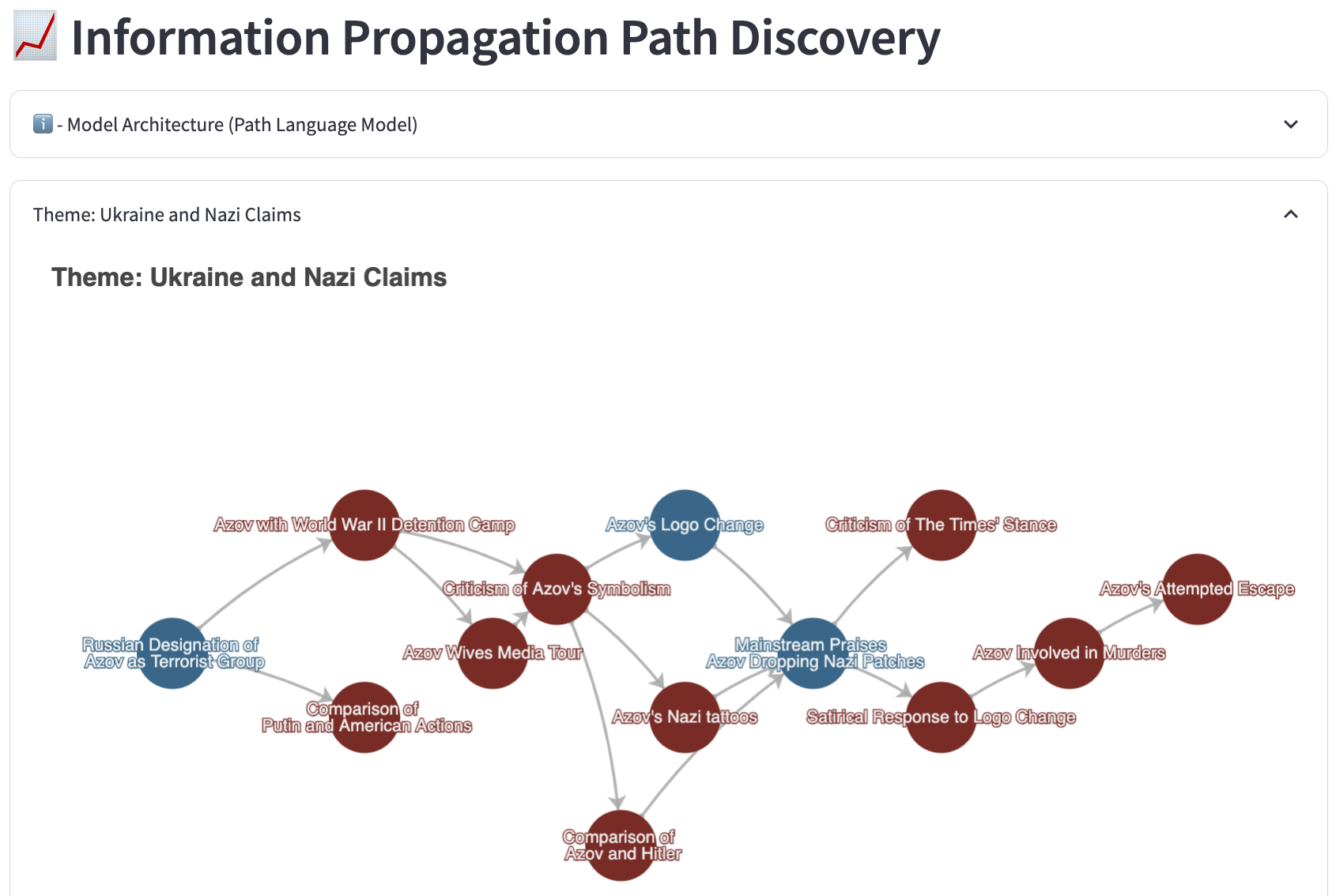}
    \caption{The interface of Information Propagation Pathway.}
    \label{summary}
    \vspace{-10pt}
\end{figure}



\subsubsection{Tweet Graph Construction.}

We experiment with Twitter data by crawling tweets with specific hashtags or topics, which are called \textit{theme} in this paper. We sample representative themes related to COVID-19, Russia-Ukraine Conflict, Philippine Tension, etc. The dataset contains 56,882 tweets over 6 months. We create edges between tweets within the same theme based on their temporal orders.  

\noindent
\textbf{Tweet Clustering.}
We perform K-means clustering \cite{macqueen1967some,Lloyd1982LeastSQ} to group tweets based on semantic similarity using Sentence-BERT \cite{reimers2019sentencebert}. 
We find an appropriate number of clusters based on Silhouette Score \cite{ROUSSEEUW198753}, which considers the cohesion and separation of the clusters.

\noindent
\textbf{Claim Summarization.}
Next, we summarize the claim (central idea) of a cluster using ChatGPT \cite{brown2020language}. The prompt is ``\textit{Summarize the central idea of the following list of tweets}''. To select high-quality tweets, we filter tweets based on the distance with the cluster center. 

\noindent
\textbf{Transition Matrix Construction.}
The final step is to build transition edges between claims. As shown in Figure 3, we  construct a transition matrix $T \in \mathbb{R}^{n \times n}$, where $n$ is the total number of claims, and $t_{ij}$ represents the total transitions from the tweet in claim cluster $c_{i}$ to the tweet in claim cluster $c_{j}$ . Next, we normalize the transition matrix by the total number of transitions, which represents the probability matrix between claims. To build the pattern graph, we first filter out the transitions with low probability, as they may not represent the general cases. After that, we view the claims as nodes, and the transitions as edges. In this way, we obtain a pattern graph representing the claims flow over time. 

\section{Conclusion and Future Work}

We present a demo for analyzing the interplay between language and human ideology. We first simulate how human frame messages with different ideology (i.e., \textit{``we shape our buildings''}). After that, we uncover information propagation patterns to reveal how claims can influence ideology communities (i.e., \textit{``thereafter they shape us''}).  
Future research includes the fine-grained community detection via modeling influence in social networks.

\bibliography{aaai24}

\end{document}